\documentclass{JHEP3}

\usepackage{Pformel,Pmeta}
\usepackage{amsmath,amsfonts,amsthm,times}

\preprint{PTA/07-46}

\title{Boundary action of the $\Hp$ model}

\author{Vladimir Fateev$^{1,2}$
and Sylvain Ribault$^1$ 
\\ \!\!\!$^1$\!
 Laboratoire de Physique Th\'eorique et Astroparticules, UMR5207 CNRS-UM2,
 \\
 Universit\'e Montpellier II, Place E. Bataillon,
 \\
 34095 Montpellier Cedex 05, France 
 \\
\!\!\!$^2$\! Landau Institute for Theoretical Physics
 \\
 142432 Chernogolovka, Russia
 \\
 {\footnotesize \tt fateev,ribault@lpta.univ-montp2.fr }
}

\abstract{We find the boundary action for Euclidean $AdS_2$ D-branes in $\Hp$. This action is consistent with the D-branes' symmetries and with the $\Hp$-Liouville relation for disc correlators. It can be used for performing free-field calculations in the $\Hp$ model with boundaries. We explicitly perform the Coulomb-like integrals which appear in the free-field calculation of the bulk one-point function, and find agreement with previously known conformal bootstrap results. }

\makeatletter\let\default@color\current@color\makeatother 

\begin{document}

\zeq\section{Introduction and summary}

The $\Hp$ model on a sphere was solved ten years ago thanks to the methods of the conformal bootstrap \cite{tes97a}, which rely on symmetry and consistency assumptions and do not exploit the Lagrangian definition of the model. It was later realized that concordant information on the structure constants of the $\Hp$ model could independently be derived thanks to the so-called free-field approach \cite{ios00,hos00,gn01}, which consists of perturbative calculations based on the Lagrangian definition. Then, after the $AdS_2$ D-branes in $AdS_3$ \cite{bp00}, Euclidean $AdS_2$ D-branes were discovered in $\Hp$ \cite{pst01,lop01}, which is the Euclidan version of $AdS_3$. The worldsheet description of strings ending on such D-branes is the $\Hp$ model on a disc with maximally symmetric boundary conditions. The solution of this model by conformal bootstrap methods was recently completed \cite{hr06,rib07}. 

These developments have left the problem of the Lagrangian definition of the $\Hp$ model on the disc open. In other words, what is the boundary action for the Euclidean $AdS_2$ D-branes? This action may be useful for obtaining a more synthetic perspective on the model, relating it to other models, and solving it on higher genus Riemann surfaces with boundaries. Path-integral calculations were indeed very helpful in the recent study of the $\Hp$ model on higher genus closed Riemann surfaces \cite{hs07}. 

The problem of finding the boundary action was addressed by Ponsot and Silva \cite{ps02}, who showed that the variations of the bulk $\Hp$ action already vanished by themselves in the presence of $AdS_2$ D-branes. They concluded that the boundary action vanished. However, $AdS_2$ D-branes come in a family with a continuous parameter $c$, and the results of path-integral calculations should depend on $c$. One could try to impose the $c$-dependent gluing conditions as constraints on the path integral, but it is not clear how to compute the resulting constrained integral.
Here, we will instead propose a boundary action (\ref{sbdy}) which vanishes on-shell but nevertheless contributes to path-integral calculations. We will show that the expected boundary conditions can be derived from this action.

We will argue that the path-integral expressions of disc correlators which follow from our boundary action agree with the known disc correlators. The argument relies on the simple relation of the known disc correlators in the $\Hp$ model with disc correlators in Liouville theory \cite{hr06}. It was recently shown that the relation between $\Hp$ and Liouville correlators on a sphere could easily be derived by a formal path-integral calculation \cite{hs07}. We will sketch a similar calculation in the case of correlators on a disc. 

We will also argue that the boundary action can be used for performing free-field calculations. We will indeed first check that the boundary action preserves the expected current-algebra symmetries. We will then  compute the bulk one-point function and find agreement with the conformal bootstrap result of \cite{pst01,lop01}. This free-field computation will involve the explicit determination of a family of  bulk-boundary Coulomb-like integrals (\ref{bigint}). We will actually only use particular cases of such integrals; the most general integrals would appear in free-field calculations of bulk-boundary two-point functions in Liouville theory and in the $\Hp$ model.

We are informed that a similar setup for the boundary dynamics of the $\Hp$ model was found by T. Creutzig and V. Schomerus in connection with their work on the $GL(1|1)$ supergroup WZNW model \cite{cs07}.

The basic concepts of boundary conformal field theory and non-rational conformal field theory which we will use are explained in the review articles \cite{sch02,sch05}.

\zeq\section{Classical analysis of the $\Hp$ model with a boundary}

Let us first define the bulk $\Hp$ model on the complex plane, which is conformally equivalent to the Riemann sphere. We will parametrize the plane with a complex variable $z=\tau+i\sigma$ and denote $\iint \equiv \int d^2z$; the single integration symbol $\int$ will be reserved for integrals over the boundary $z=\bz$ of the upper half-plane. The model is defined by the following action, where we adopt the notations of \cite{hs07} (while adding a ``bulk cosmological constant'' numerical factor $\lambda$ to the interaction term)
\bea
S^{bulk}=\frac{1}{2\pi} \iint \left(\p\phi \bp\phi + \beta\bp \g + \bar{\beta}\p\bg -\lambda b^2 \beta\bar{\beta} e^{2b\phi}\right)\ .
\label{sbulk}
\eea
All fields are bosonic. The field $\g$ has conformal dimension zero, the holomorphic field $\beta$ has conformal dimension one, and $\phi$ has conformal dimension zero but a background charge $b$, so that the interaction term $\iint \beta\bar{\beta} e^{2b\phi}$ is conformally invariant with respect to the holomorphic stress-energy tensor 
\bea
T=-\beta\p\g-\p\phi^2+b\p^2\phi\ .
\eea 
Here $b>0$ is a continuous parameter of the $\Hp$ model, which is related to the level $k$ and the central charge $c$ by
\bea
b^{2} = \frac{1}{k-2} \ , \ c=\frac{3k}{k-2}\ .
\eea

The model actually has not only a conformal symmetry, but also an affine symmetry generated by currents which we denote as
\bea
\begin{array}{rclcrcl}
J^-&=& \beta\ ,  &\hspace{4mm}  \hspace{4mm} & \bJ^- &=& \bar{\beta}\ ,
\\
J^3&=& \beta\gamma+b^{-1}\p\phi\ , &   & \bJ^3 &=& \bar{\beta}\bg +b^{-1}\bp\phi \ ,
\\
J^+&=& \beta\gamma^2 +2b^{-1}\gamma\p\phi -k\p\gamma\ , &  & \bJ^+ &=&  \bar{\beta}\bg^2+2b^{-1}\bg\bp\phi-k\bp\bg\ .
\end{array}
\label{jbg}
\eea
Euclidean $AdS_2$ D-branes are maximally symmetric in that they preserve half of these six currents. Strings ending on such D-branes are described by the $\Hp$ model on the complex upper half-plane, with 
the following gluing conditions at $z=\bz$ \cite{pst01,ps02}:
\bea
\begin{array}{rcl}
 J^- + \bJ^- &=& 0 \ , \\ J^3-\bJ^3 &=& 0 \ , \\ J^++\bJ^+ &=& 0 \ .
\end{array}
\label{goc}
\eea
There is actually a one-parameter family of D-branes which satisfy such gluing conditions; namely, for any real number $c$, we can assume that at $z=\bz$
\bea
\begin{array}{rcl}
 \beta+\bar{\beta} & = & 0 \ , \\ \g+\bg &=& ce^{b\phi}\ , \\ (\bp-\p) \phi &=& cb\beta e^{b\phi}\ .
\end{array}
\label{gof}
\eea
To these gluing conditions, one may add the bulk equations of motion for the $\beta,\bar{\beta}$ fields, which have no reason to fail at the boundary $z=\bz$:
\bea
\p\bg = \lambda b^2\beta e^{2b\phi} \ \ , \ \ \bp \g = \lambda b^2 \bar{\beta} e^{2b\phi} \ .
\eea
Modulo these equations, our gluing conditions are equivalent to the known gluing conditions on $(\phi,\g,\bg)$ \cite{ps02}. 

Let us clarify a subtlety about the derivation of the gluing conditions (\ref{goc}) for the currents (\ref{jbg}) from the gluing conditions (\ref{gof}) for the fields $(\phi,\beta,\gamma)$. This derivation should take into account the nature of $(\phi,\beta,\gamma)$ as quantum fields, so that the currents (\ref{jbg}) involve regularized products of these fields. The regularization involves the explicit subtraction of the singularities in the operator products; these singularities can be deduced from eq. (\ref{cont}). Alternatively, we could do the same calculation in a classical framework which would treat the fields as ordinary functions on the worldsheet; but in this framework the expressions for the currents $J^+,\bJ^+$ differ from our formula (\ref{jbg}). The classical version of $J^+$ is indeed $J^+_{cl}=\beta\g^2+2b^{-1}\g\p\phi-(k-2)\p\g$, where the difference $J^+_{cl}-J^+=2\p\g$ arises from regularizing the operator product $\beta\g^2$.\footnote{We are very grateful to the JHEP referee for helping us clarify this point.}

Now consider the bulk action $S^{bulk}$ (\ref{sbulk}) on the upper half-plane. Cancelling the variations of this action implies the bulk equations of motion, plus some constraints on the behaviour of the fields at the boundary $z=\bz$. It was found in \cite{ps02} that these constraints are satisfied by the gluing conditions (\ref{gof}) for all values of $c$. (The calculations in \cite{ps02} were actually performed using the equivalent action obtained by integrating out the fields $\beta,\bar{\beta}$ in the path integral.) However, the bulk action by itself cannot be enough for defining the quantum dynamics of the model, because it does not know the value of $c$. It is however still possible to add a boundary term to the action provided it vanishes when the gluing conditions (\ref{gof}) are obeyed, and we propose
\bea
S^{bdy} = \frac{i}{4\pi} \int \beta \left(\g+\bg -c e^{b\phi}\right) \  ,
\ \ \ \ \ \beta+\bar{\beta}\ \underset{z\rar\bz}{=}0\ ,
\label{sbdy}
\eea
where the single integral $\int=\int d\tau$ means the integral over the boundary $z=\bz$ of the upper half-plane. We therefore propose that in path-integral calculations the first gluing condition is imposed as a constraint, while the last two should follow from the variational principle applied to the action $S=S^{bulk}+S^{bdy}$. 

Let us now study the variations of the action. Using an integration by parts and $\p=\frac12(\p_\tau-i\p_\sigma)$, the action is rewritten as
\bea
S=S^{bulk}+S^{bdy} = \frac{1}{2\pi} \iint \left(\p\phi \bp\phi - \g\bp\beta - \bg\p\bar{\beta} -\lambda b^2 \beta\bar{\beta} e^{2b\phi}\right) - c\frac{i}{4\pi} \int \beta e^{b\phi}\ .
\label{stot}
\eea
Taking into account the constraint $\delta(\beta+\bar{\beta})\underset{z\rar\bz}{=}0$, the boundary terms in the variations of the action are therefore
\bea
(\delta S)^{bdy} = \frac{i}{4\pi} \int \left[-\delta\phi\left((\p-\bp)\phi + cb\beta e^{b\phi}\right) +\delta\beta\left(\g + \bg-ce^{b\phi}\right)\right] \ .
\eea
Requiring the vanishing of the coefficients of the independent variations $\delta\phi$ and $\delta\beta$ therefore yields the last two gluing conditions in eq. (\ref{gof}). This shows that our proposal for the boundary action is classically sound. Not only it does not spoil the compatibility of the desired gluing conditions with the variational principle, but also it singles out a value for the parameter $c$. 

\zeq\section{Path-integral derivation of the relation with Liouville theory}

Let us now consider the path-integral representation of a general $\Hp$ correlator on the upper half-plane (which is conformally equivalent to the disc), with a number of bulk and boundary operator insertions. We will show how this correlator is related to a Liouville theory correlator by integrating out the fields $\g,\bg$ and then $\beta,\bar{\beta}$. The calculation follows closely that of Hikida and Schomerus in the case of the sphere \cite{hs07}, so we will only sketch the few most relevant points.

The relevant bulk and boundary operators, with spins $j$ and $\ell$, isospins $\mu$ and $\nu$ and worldsheet positions $z$ and $\tau$ respectively, are
\bea
\Phi^{j}(\mu|z) &=& |\mu|^{2j+2} e^{\mu\g(z)-\bar{\mu}\bg(\bz)} e^{2b(j+1)\phi(z,\bz)}\ , 
\label{pjm}
\\ \Psi^{\ell}(\nu|\tau) &=& |\nu|^{\ell+1} e^{\frac12\nu(\g(\tau)-\bg(\tau))} e^{b(\ell+1)\phi(\tau)}\ .
\label{pln}
\eea
The correlator to be computed is
\bea
\Omega =  \int {\cal D}\phi\ {\cal D}\beta\ {\cal D}\bar{\beta}\ {\cal D}\g\ {\cal D}\bg\ e^{-S}\ \prod_{i=1}^n \Phi^{j_i}(\mu_i|z_i)\ \prod_{a=1}^m\Psi^{\ell_a}(\nu_a|\tau_a)\ ,
\eea
where the constraint $\beta+\bar{\beta}\underset{z\rar\bz}{=}0$ is implicitly understood, and the action $S$ is given in eq. (\ref{stot}), where the boundary parameter $c$ can actually jump at the insertion points of boundary operators. Note that the sign of the action is such that the Gaussian integral over $\beta,\bar{\beta}$ is convergent provided $\beta$ and $\bar{\beta}$ are complex anticonjugates. Integrating out $\beta,\bar{\beta}$ would produce the well-known $\Hp$ sigma model \cite{gaw91}, plus an extra boundary action.

As an aside, recall that large $k$ limits of $\Hp$ correlators can then be determined thanks to so-called minisuperspace computations. In such computations, the functional integrals $\int {\cal D}\phi\ {\cal D}\g\ {\cal D}\bg $ are replaced with ordinary integrals over the zero-modes $\int d\phi\ d^2\g$. Due to factors $\p\bg$ or $\bp\g$, the bulk and boundary interaction terms then vanish. This provides an a posteriori justification for the minisuperspace calculations of the bulk one-point function \cite{pst01}, bulk-boundary two-point function \cite{hr06}, and boundary three-point function \cite{rib07}, which did not involve any contributions from the then-unknown boundary action. (The results of such minisuperspace calculations nevertheless depend on the boundary parameter $c$ because the zero-modes have to be integrated only over the D-brane's world-volume.) 

Let us perform the path integral over $\g,\bg$. This yields 
delta-function constraints on derivatives of the fields $\beta,\bar{\beta}$, namely
\begin{multline}
\Omega = \int {\cal D}\phi\ {\cal D}\beta\ {\cal D}\bar{\beta}\ \ \delta\left(\frac{1}{2\pi}\bp\beta(z)-\sum_i\mu_i\delta^{(2)}(z-z_i)-\frac12\sum_a \nu_a\delta^{(2)}(z-\tau_a)\right)\\
\times \ \delta\left(\frac{1}{2\pi}\p\bar{\beta}(\bz)+\sum_i\bar{\mu}_i\delta^{(2)}(z-\bz_i) +\frac12 \sum_a\nu_a\delta^{(2)}(z-\tau_a)\right)\  \times \cdots\ .
\end{multline}
Now, performing the path integral over $\beta,\bar{\beta}$ (subject to $\beta+\bar{\beta}\underset{z\rar\bz}{=}0$) will yield a nonzero result only provided $\sum_i(\mu_i+\bar{\mu_i})+\sum_a\nu_a=0$, and will force $\beta,\bar{\beta}$ to adopt the values
\bea
\beta_s(z) &=& \sum_i\frac{\mu_i}{z-z_i} + \sum_i\frac{\bar{\mu}_i}{z-\bz_i} +\sum_a\frac{\nu_a}{z-\tau_a}\ ,
\\
\bar{\beta}_s(\bz) &=& -\sum_i\frac{\mu_i}{\bz-z_i} - \sum_i\frac{\bar{\mu}_i}{\bz-\bz_i} -\sum_a\frac{\nu_a}{\bz-\tau_a}\ .
\eea
Notice that the $\sum_i\frac{\bar{\mu}_i}{z-\bz_i}$ terms of $\beta(z)$ do not contribute to $\bp\beta(z)$, which is defined only in the upper half-plane whereas $\bz_i$ belong to the lower half-plane. However, such terms are required by the assumed condition $\beta+\bar{\beta}\underset{z\rar\bz}{=}0$. Note also the subtlety in defining $\delta^{(2)}(z-\tau_a)$ when $\tau_a$ belongs to the boundary; the correct treatment of this subtlety (for instance by slightly moving $\tau_a$ into the upper half-plane) leads to the correct numerical factor of the term $\sum_a\frac{\nu_a}{z-\tau_a}$ in $\beta_s(z)$. 

After replacing the fields $\beta,\bar{\beta}$ by their values $\beta_s,\bar{\beta}_s$ in the path integral,
we should relate $\phi$ to the Liouville field so that $\Omega$ can be interpreted as a Liouville theory correlator (plus some simple factors). This is achieved 
by performing a change of variable on the field $\phi$ so that $\beta_s\bar{\beta}_s e^{2b\phi}= - e^{2b\tilde{\phi}}$, where $\tilde{\phi}$ is the Liouville field. In the case of the sphere, the effect of this change of variable on the kinetic term $\iint \p\phi\bp\phi$ could be interpreted as the introduction of degenerate Liouville operators at the zeroes of $\beta_s$, and the situation is the same in our case of the disc. We refer to \cite{hs07} for the details. The new feature in our case is the presence of the boundary term $-c\frac{i}{4\pi}\int \beta_s e^{b\phi}$. The change of variable $\phi\rar \tilde{\phi}$ will only absorb $\beta_s$ into the exponential up to an overall sign:
\bea
 -c\frac{i}{4\pi}\int \beta_s e^{b\phi} = -c\frac{i}{4\pi}\int (\sgn\beta_s)\ e^{b\tilde{\phi}} \ .
 \eea
The value of the Liouville boundary cosmological constant (i.e. of the coefficient of $\int e^{b\tilde{\phi}}$) is therefore
 \bea
\mu_B = -c\frac{i}{4\pi}  \sgn\beta_s 
 \label{mub}
 \eea
This relation between the $\Hp$ and Liouville boundary parameters, and the rest of the $\Hp$-Liouville relation on the disc whose derivation we just sketched, fully agree with the known $\Hp$-Liouville relation on the disc \cite{hr06}, which was originally derived by conformal bootstrap methods. 
In particular, $\mu_B$ is pure imaginary for physical (i.e. real) values of $c$, and its sign is determined by the sign of $\beta_s$. 
This agreement amounts to an additional heuristic argument in favour of our boundary action $S^{bdy}$ (\ref{sbdy}).

\zeq\section{Free-field formalism}

The total action $S$ (\ref{stot}) can be split into free terms, plus bulk and boundary interaction terms corresponding to the Lagrangians
\bea
L^{bulk}=\beta\bar{\beta}e^{2b\phi}, \  L^{bdy}=\beta e^{b\phi}\ .
\eea
The free theory is subject to the simple gluing conditions (which coincide with the gluing conditions (\ref{gof}) at $c=0$)
\bea
\begin{array}{rcl}
\beta+\bar{\beta}&=&0\ ,
\\
\g +\bg &=& 0\ ,
\\
(\p-\bp) \phi &=& 0\ .
\end{array}
\eea
The non-vanishing pairings of the basic fields in the presence of such gluing conditions are
\bea
\begin{array}{rcl}
\la \phi(z)\phi(w) \ra &=&-\log|z-w||\bz-w|\ ,
\\
\la \beta(z)\gamma(w) \ra &=& \frac{1}{w-z}\ ,
\\
\la \bar{\beta}(\bz)\g(w) \ra &=& -\frac{1}{w-\bz}\ ,
\\
\la \beta(z)\bg(\bar{w}) \ra &=& -\frac{1}{\bar{w}-z}\ ,
\\
\la \bar{\beta}(\bz)\bg(\bar{w}) \ra &=& \frac{1}{\bar{w}-\bz}\ . 
\end{array}
\label{cont}
\eea
These correlators are consistent with the fields $\beta,\g$ being holomorphic, and the fields $\bar{\beta},\bg$ being antiholomorphic, as implied by their respective bulk equations of motion in the free theory. They also agree with the gluing conditions, in the sense that
\bea
\underset{z\rar \bz}{\lim} \la (\beta(z)+\bar{\beta}(\bz)) \cdots \ra = \underset{z\rar \bz}{\lim} \la (\g(z)+\bg(\bz)) \cdots \ra = 0\ .
\eea
(For most purposes, the $\beta\g$ system with conformal weights $(1,0)$ is actually equivalent to a suitably normalized complex free boson $\omega$ such that $\beta=\p\omega,\bar{\beta}=\bp\omega,\g=\omega_L^*,\bg=\omega_R^*$, where $\omega_{L},\omega_{R}$ are the holomorphic and antiholomorphic terms of $\omega$ respectively, and the star denotes complex conjugation.)

Let us check that the bulk and boundary interaction terms preserve the affine symmetries (\ref{goc}). Our treatment of this problem is inspired from \cite{fgk07}, where more general results and references on perturbed boundary conformal field theories can be found. To first order in the boundary coupling constant $c$, the $J^++\bJ^+$ symmetry condition is
\bea
\underset{z\rar\bz}{\lim} \la (J^+(z)+\bJ^+(\bz)) \int L^{bdy}(\tau) \cdots \ra = 0\ .
\label{bic}
\eea
Notice that $\underset{z\rar\bz}{\lim} \la (J^+(z)+\bJ^+(\bz)) L^{bdy}(\tau) \cdots \ra$ vanishes for all $\tau$ due to the symmetries of the free theory. The symmetry condition we just wrote might nevertheless fail because of the singularities which appear as the operators $J^++\bJ^+$ and $L^{bdy}$ come close, and which might prevent the integration over $\tau$ from commuting with the limit $\underset{z\rar\bz}{\lim}$. 
Using the contractions (\ref{cont}), the singular terms coming from $J^+$ are
\footnote{Notice that $\bJ^+(\bz)L^{bdy}(\tau)\sim -\frac{b^{-2}}{(\bz-\tau)^2} e^{b\phi(\tau)} -\frac{2b^{-1}}{\bz-\tau} \p\phi(\tau)e^{b\phi(\tau)}$, which is in accordance with  the vanishing of $\underset{z\rar\bz}{\lim} \la (J^+(z)+\bJ^+(\bz)) L^{bdy}(\tau) \cdots \ra$ for any given $\tau$.}
\bea
J^+(z)L^{bdy}(\tau)\sim \frac{b^{-2}}{(z-\tau)^2} e^{b\phi(\tau)} +\frac{2b^{-1}}{z-\tau} \p\phi(\tau)e^{b\phi(\tau)} = b^{-2}\p_\tau \frac{e^{b\phi(\tau)}}{z-\tau} -\frac{i}{z-\tau} \p_\sigma\phi(\tau) e^{b\phi(\tau)}\ .
\eea
The first term is a total $\tau$-derivative and will therefore not contribute to the $\tau$-integral. The second term vanishes due to the Neumann gluing condition on $\phi$, namely $\p_\sigma\phi=0$. Similar calculations show that the singular terms coming from $\bJ^+$ also vanish, and therefore the symmetry condition (\ref{bic}) holds. The $J^3-\bJ^3$ and $J^-+\bJ^-$ symmetry conditions can similarly be checked.

Of course, this is not enough for fully establishing the current symmetries of the theory. One would need to check that say $\underset{z\rar\bz}{\lim} (J^+(z)+\bJ^+(\bz))$ vanishes when inserted into correlators with arbitrary numbers of insertions of both the bulk and the boundary interaction terms $\iint L^{bulk},\ \int L^{bdy}$. The case with just one bulk interaction term can be treated along the same lines as above, but is a bit more tedious. We abstain from displaying such calculations, because our main aim is only to check the correctness of the boundary action.

\section{Free-field calculation of the bulk one-point function}

Let us demonstrate the validity of the free-field formalism by computing the bulk one-point function. The calculation is quite similar to the free-field calculation of the bulk one-point function in Liouville theory with Neumann boundary conditions, which was sketched in \cite{fzz00}. The new features of the $\Hp$ case are the contribution of the $\beta\g$ system, and the resulting dependence of the one-point function on the isospin $\mu$ of the bulk field. In this section we will explicitly give the values of the relevant integrals over the worldsheet positions of the bulk and boundary interaction terms in the $\Hp$ model Lagrangian, which might also be useful for other applications. 

The one-point function of a bulk field $\Phi^j(\mu|z)$ (\ref{pjm}) in the presence of an $AdS_2$ D-brane with parameter $c$ is expected to be amenable to a free-field calculation only for certain quantized values of the spin $j$, namely
\bea
2j+1=-n\in -\N\ .
\label{jquant}
\eea
The one-point function actually has simple poles at such values of the spin $j$, whose residues are expressed as
\begin{multline}
\underset{2j+1=-n}{\rm Res}
\la \Phi^j(\mu|z) \ra_c^{\Hp} = \frac{1}{2b} |\mu|^{2j+2} \sum_{\begin{smallmatrix} m,\ell=0\\ 2m+\ell=n\end{smallmatrix}}^\infty \frac{1}{m!\ell !} 
\prod_{i=1}^m \iint d^2w_i \prod_{k=1}^\ell \int dx_k
\\
\la e^{\mu\g(z)-\bar{\mu}\bg(\bz)} e^{2b(j+1)\phi(z,\bz)} \prod_{i=1}^m \lambda
\frac{b^2}{2\pi}\beta\bar{\beta} e^{2b\phi}(w_i) \prod_{k=1}^\ell \frac{ic}{4\pi} \beta e^{b\phi}(x_k) \ra\ ,
\label{onept}
\end{multline}
where the correlator on the second line is computed in the free theory described in the previous section. 

This correlator factorizes into independent $\beta\g$ and $\phi$ correlators. Remembering that the field $\phi$ has a background charge $b$, the $\phi$ correlator is non-vanishing provided
$2j+1+2m+\ell = 0$; this is the origin of the condition (\ref{jquant}). 
The $\phi$ correlator is then
\begin{multline}
\la e^{2b(j+1)\phi(iy)} \prod_{i=1}^m e^{2b\phi(w_i)}  \prod_{k=1}^\ell e^{b\phi(x_k)} \ra
=  
\left[ \prod_{k=1}^\ell (y^2+x_k^2) \prod_{i=1}^m|y^2+w_i^2|^2\right]^{b^2(n-1)} 
\\ \times |2y|^{-\frac{b^2}{2}(n-1)^2} \times
\left[\prod_{i,k}|w_i-x_k|^2\prod_{i<i'}|w_i-w_{i'}|^2\prod_{i,i'}|w_i-\bar{w}_{i'}| \prod_{k<k'}|x_k-x_{k'}|\right]^{-2b^2}\ .
\label{phifact}
\end{multline}
In the $\beta\g$ correlator, integrating over the zero-modes of the field $\g$ yields a factor $2\pi\delta(\mu+\bar{\mu})$. The rest of the computation follows from the contractions (\ref{cont}), which in particular imply
\bea
\la e^{\mu\g(z)-\bar{\mu}\bg(\bz)} \beta(w) \ra = \frac{\mu}{w-z}+\frac{\bar{\mu}}{w-\bz} = \frac{\mu(z-\bz)}{(w-z)(w-\bz)}\ .
\eea
The actual correlator is a product of such factors,
\begin{multline}
\la e^{\mu\g(iy)-\bar{\mu}\bg(-iy)} \prod_{i=1}^m \lambda\frac{b^2}{2\pi}\beta\bar{\beta}(w_i) \prod_{k=1}^\ell \frac{ic}{4\pi}\beta(x_k) \ra 
\\
= 2\pi \delta(\mu+\bar{\mu}) (-)^m \left(\lambda\frac{b^2}{2\pi}\right)^m |2y\mu|^n \left(-\frac{ic}{4\pi} \sgn\Im\mu\right)^\ell\ \prod_{i=1}^m\frac{1}{|y^2+w_i^2|^2} \prod_{k=1}^\ell \frac{1}{y^2+x_k^2}\ .
\label{bgfact}
\end{multline}
It is already clear that, after combining the factors (\ref{phifact}) and (\ref{bgfact}), the remainder of the calculation is identical to the free-field calculation  of the one-point function $\la e^{2\al\phi(iy)}\ra_{\mu_B}^{Liouville}$ of a Liouville field with parameter $\al = b(j+1)+\frac{1}{2b}$, in the presence of a Neumann boundary with cosmological constant $\mu_B=c\frac{i}{4\pi} \sgn\Im\mu$. The integral in eq. (\ref{onept}) amounts to taking the special value $a=1+b^2-b^2n$ in the following integral:
\begin{multline}
{\cal J}_{n,m}(a|y) =\frac{1}{m!(n-2m)!} \iint \prod_{i=1}^m \frac{d^2w_i}{|y^2+w_i^2|^{2a}} \int \prod_{k=1}^{n-2m} \frac{dx_k}{(y^2+x_k^2)^{a}} 
\\
\left[\prod_{i,k} |w_i-x_k|^2 \prod_{i<i'}|w_i-w_{i'}|^2 \prod_{i,i'} |w_i-\bar{w}_{i'}|\prod_{k<k'} |x_k-x_{k'}| \right]^{-2b^2} \ ,
\label{bigint}
\end{multline}
where as before the double integrals are over the upper half-plane, and the single integrals over the real line.
This integral can be evaluated explicitly. With the notations $s(x)\equiv \sin \pi x$ and $C^i_n=\frac{n!}{i!(n-i)!}$, the result is
\bea
{\cal J}_{n,m}(a|y) &=& \frac{|2y|^{n(1-2a-(n-1)b^2)}}{n!} \left(\frac{2\pi}{\G(1-b^2)}\right)^n \frac{2^{-2m}}{(s(b^2))^m} I_n(a) J_{n,m}(a) \ ,
\eea
where
\bea
I_n(a) &=& \prod_{i=0}^{n-1} \frac{\G(1-(i+1)b^2) \G(2a-1+(n-1+i)b^2)}{\G^2(a+ib^2)}\ ,
\eea
and
\bea
J_{n,m}(a) &=& \sum_{i=0}^m (-)^i C_{n-m-i}^{m-i} \frac{s((n+1-2i)b^2)}{s((n+1-i)b^2)} \prod_{r=0}^{i-1}\frac{s((n-r)b^2) s(a+(n-r)b^2)}{s((r+1)b^2)s(a+rb^2)}\ .
\eea

Let us introduce new notations for the boundary parameter $c$:
\bea
&& c=-\sqrt{\lambda\frac{8\pi b^2}{\sin\pi b^2}} \sinh r=-i\sgn\Im\mu\ \sqrt{\lambda\frac{8\pi b^2}{\sin\pi b^2}} \cosh\sigma \ ,
\\ &&
{\rm where}\ \ \ \ \sigma=r-i\frac{\pi}{2}\sgn\Im\mu\ .
\eea 
We now sum over the numbers of screening charges $m,\ell$ while keeping $2m+\ell=n$ fixed like in eq. (\ref{onept}). This summation reduces to the following formula, which can be proved by the application of standard trigonometric identities: 
\bea
\sum_{m=0}^{[n/2]}(-)^m(2\cosh \sigma)^{n-2m}J_{n,m}(a)=\sum_{i=0}^n\cosh((n-2i)\sigma)\frac{[-nb^2]_i [-(n-1)b^2-a]_i}{[b^2]_i [a]_i}\ ,
\label{sj}
\eea
where we use the notation $[x]_i\equiv \prod_{r=0}^{i-1} s(x+rb^2)$.


Actually, our calculation involves a relatively simple case of these formulas, because it only uses  particular values of the parameter $a$, namely the values $a=1+b^2-b^2n$ with $n\in\N$. In this case, only the term $i=0$ survives in eq. (\ref{sj}). The bulk-boundary two-point functions in Liouville theory and in $\Hp$ would involve the general case. And the integral (\ref{bigint}) might be useful for other applications.

The result of the free-field calculation is therefore:
\begin{multline}
\underset{2j+1=-n}{\rm Res} \la \Phi^j(\mu|iy) \ra_{c}^{\Hp} =|2y|^{\frac{b^2}{2}(n^2-1)} \delta(\mu+\bar{\mu}) |\mu| \times  \pi b^{-1} \left(\frac{2}{\lambda}\frac{\G(1-b^2)}{\G(1+b^2)}\right)^{-\frac{n}{2}}
\\ \times
\frac{(-)^n}{n!} \G(1-b^2n) \cosh n \left(r-i\tfrac{\pi}{2}\sgn\Im\mu\right)\ ,
\end{multline}
This should be compared to the conformal bootstrap result \cite{pst01,lop01}, which in the notations of \cite{rib05b} reads: 
\begin{multline}
\la \Phi^j(\mu|iy) \ra_{c}^{\Hp} = |2y|^{2b^2j(j+1)} \delta(\mu+\bar{\mu}) |\mu| 
\times
\pi (8b^2)^{-\frac14} \left(\pi\frac{\G(1-b^2)}{\G(1+b^2)}\right)^{j+\frac12}
\\ \times
\G(2j+1) \G(1+b^2(2j+1)) \cosh (2j+1)\left(r-i\tfrac{\pi}{2}\sgn\Im\mu\right)\ .
\label{bres}
\end{multline}
We thus find agreement, provided the bulk cosmological constant is chosen as $\lambda=\frac{2}{\pi}$, so that the normalizations of the bulk interaction term agree. The remaining discrepancy is just an
overall numerical factor. The numerical normalization factor in eq. (\ref{bres}) was derived in \cite{pst01} by a ``modular bootstrap'' calculation of the annulus amplitude, whereas we did not impose such a normalization here.

This agreement provides another test of the proposed boundary action of the $\Hp$ model.

\acknowledgments{We are grateful to Stefan Fredenhagen, Volker Schomerus and Thomas Creutzig for helpful discussions and correspondence. SR would like to thank the organizers of the August 2007 Ascona conference on the symmetries of String Theory for a good working atmosphere.}


\begin{thebibliography}{10}
\expandafter\ifx\csname url\endcsname\relax
  \def\url#1{{\tt #1}}\fi
\expandafter\ifx\csname urlprefix\endcsname\relax\def\urlprefix{URL }\fi
\providecommand{\eprint}[2][]{\url{#2}}

\bibitem{tes97a}
J.~Teschner, {\em On structure constants and fusion rules in the
  $SL(2,\mathbb{C})/SU(2)$ {WZNW} model\/}, Nucl. Phys. B546 pp. 390--422
  (1999), \eprint{hep-th/9712256}

\bibitem{ios00}
N.~Ishibashi, K.~Okuyama, Y.~Satoh, {\em Path integral approach to string
  theory on AdS(3)\/}, Nucl. Phys. B588 pp. 149--177 (2000),
  \eprint{hep-th/0005152}

\bibitem{hos00}
K.~Hosomichi, K.~Okuyama, Y.~Satoh, {\em Free field approach to string theory
  on AdS(3)\/}, Nucl. Phys. B598 pp. 451--466 (2001), \eprint{hep-th/0009107}

\bibitem{gn01}
G.~Giribet, C.~Nunez, {\em Correlators in AdS(3) string theory\/}, JHEP 06 p.
  010 (2001), \eprint{hep-th/0105200}

\bibitem{bp00}
C.~Bachas, M.~Petropoulos, {\em {Anti-de-Sitter D-branes}\/}, JHEP 02 p. 025
  (2001), \eprint{hep-th/0012234}

\bibitem{pst01}
B.~Ponsot, V.~Schomerus, J.~Teschner, {\em Branes in the {Euclidean
  $AdS_3$}\/}, JHEP 02 p. 016 (2002), \eprint{hep-th/0112198}

\bibitem{lop01}
P.~Lee, H.~Ooguri, J.-w. Park, {\em Boundary states for {AdS(2) branes in
  AdS(3)}\/}, Nucl. Phys. B632 pp. 283--302 (2002), \eprint{hep-th/0112188}

\bibitem{hr06}
K.~Hosomichi, S.~Ribault, {\em Solution of the $H_3^+$ model on a disc\/}, JHEP
  01 p. 057 (2007), \eprint{hep-th/0610117}

\bibitem{rib07}
S.~Ribault, {\em Boundary three-point function on AdS2 D-branes\/}  (2007),
  \eprint{arXiv:0708.3028 [hep-th]}

\bibitem{hs07}
Y.~Hikida, V.~Schomerus, {\em $H^+_3$ WZNW model from Liouville field theory\/}
   (2007), \eprint{arXiv:0706.1030 [hep-th]}

\bibitem{ps02}
B.~Ponsot, S.~Silva, {\em Are there really any {AdS(2)} branes in the euclidean
  (or not) {AdS(3)}?\/}, Phys. Lett. B551 pp. 173--177 (2003),
  \eprint{hep-th/0209084}

\bibitem{cs07}
T.~Creutzig, V.~Schomerus, {\em \rm work in progress\/}

\bibitem{sch02}
V.~Schomerus, {\em Lectures on branes in curved backgrounds\/}, Class. Quant.
  Grav. 19 pp. 5781--5847 (2002), \eprint{hep-th/0209241}

\bibitem{sch05}
V.~Schomerus, {\em Non-compact string backgrounds and non-rational CFT\/},
  Phys. Rept. 431 pp. 39--86 (2006), \eprint{hep-th/0509155}

\bibitem{gaw91}
K.~Gawedzki, {\em Noncompact WZW conformal field theories\/}  (1991),
  \eprint{hep-th/9110076}

\bibitem{fgk07}
S.~Fredenhagen, M.~R. Gaberdiel, C.~A. Keller, {\em Symmetries of perturbed
  conformal field theories\/}  (2007), \eprint{arXiv:0707.2511 [hep-th]}

\bibitem{fzz00}
V.~Fateev, A.~B. Zamolodchikov, A.~B. Zamolodchikov, {\em Boundary Liouville
  field theory. I: Boundary state and boundary two-point function\/}  (2000),
  \eprint{hep-th/0001012}

\bibitem{rib05b}
S.~Ribault, {\em Discrete D-branes in $AdS_3$ and in the 2d black hole\/}, JHEP
  08 p. 015 (2006), \eprint{hep-th/0512238}

\end{thebibliography}

\end{document}